# Spin-dependent transport through magnetic nanojunctions

Kamil Walczak[1,2] and Gloria Platero[2]

[1] Institute of Physics, A. Mickiewicz University, Umultowska 85, 61-614 Poznań, Poland
[2] Instituto de Ciencia de Materiales de Madrid, CSIC, Cantoblanco, 28049 Madrid, Spain

Coherent electronic transport through a molecular device is studied using non-equilibrium Green's function (NEGF) formalism. Such device is made of atomic nanowire which is connected to ferromagnetic electrodes. The molecule itself is described with the help of Hubbard model (Coulomb interactions are treated by means of the Hartree-Fock approximation), while the coupling to the electrodes is modeled through the use of a broad-band theory. It was shown that magnetoresistance varies periodically with increasing length of the atomic wire (in the linear response regime) and oscillates with increasing bias voltage (in the nonlinear response regime). Since the TMR effect for analyzed structures is predicted to be large (tens of percent), these junctions seem to be suitable for application as magnetoresistive elements in future electronic circuits.

## I. Introduction

Advances in experimental techniques have made it possible to fabricate molecular-scale devices and to measure their current-voltage (I-V) characteristics. Generally speaking, molecular junctions are made of two (or more) electrodes connected by a molecule (or molecular layer). Experiments performed on such structures have shown a variety of transport phenomena: rectification [1], negative differential resistance [2], switching behavior, memory cell operation [3], and transistor action [4]. Molecular junctions are important both from a pure science point of view and because of their potential applications. They are promising candidates for future electronic devices because of their small sizes and the theoretically inexhaustible structural modifications of the molecules. They also have the potential to become relatively cheap and easily obtained layer-based molecular junctions (due to self-assembly features of organic molecules). Transport properties of such devices are dominated mainly by effects such as: quantization of molecular energy levels and discreteness of electron charge. However, it should be also stressed that an electron's spin as well as its charge can be employed to store, process, and transmit information [5, 6]. Since the spin orientation of conduction electrons survives for a long period of time ($\sim ns$) in comparison with the residence time of the traveling electron on the molecules ($\sim fs$), molecular junctions may be useful in applications involving electron spin manipulation. It means that spin-conserving transport in molecular-scale devices is possible, where spin-flip scattering and spin-orbit processes can be neglected.

Recent experiments on Ni nanocontacts disclosed magnetoresistance values up to 280% at room temperature for a few-atom contact in the ballistic transport regime [7]. This effect is explained by a relative change of a number of conducting channels when magnetization changes its orientation from parallel to antiparallel [8]. Similar effects are also expected for molecular junctions, where single molecular wires are attached to ferromagnetic electrodes [9-13]. The relative orientations of the electrode magnetizations can be changed from parallel (P) to antiparallel (AP) by applying an external magnetic field. In this case, tunnel magnetoresistance (TMR) defined as a relative difference of conductances in the P and AP alignments is associated with asymmetry of a density of states (DOS) for two spin channels in ferromagnetic materials. Furthermore, the TMR effect is influenced by the following factors: the electronic structure of nanowire, the nature of molecule-to-electrodes coupling and the location of the Fermi level in relation to molecular energy levels. In particular, spin-polarized transport of electrons flowing through the junction consisting of a self-assembled monolayer (SAM) of octanethiol attached to a pair of Ni electrodes was studied experimentally [14]. These molecular junctions exhibit TMR values up to 16% at low bias voltages. However, strong voltage and temperature dependence of the junction magnetoresistance and time-dependent telegraph noise signals suggest that transport properties of this device can be also affected by localized states in the molecular monolayer.

The main purpose of the present work is to study the coherent spin-dependent electronic transport through atomic nanowires symmetrically coupled to a pair of identical ferromagnetic electrodes. This choice is dictated by the experimental situation, in which linear carbon-atom chains containing up to 20 atoms connected at the ends to metal atoms have been synthesized [15] and subsequently recognized as ideal one-dimensional wires [16,17]. For the case of paramagnetic electrodes, previous *ab initio* studies show that linear conductance of carbon nanowires varies in an oscillatory manner as the number of carbon atoms is increased [16,17]. Odd-number carbon atom wires show a higher conductance than even-number ones, where the conductance values fall in the approximate range of $G_0$ to $2G_0$. Here $G_0 \equiv 2e^2/h \approx 77.5$ μS is a quantum of conductance corresponding to a resistance of 12.9 kΩ. Furthermore, an oscillatory behavior has been observed in other properties of carbon nanowires, such as their stability towards fragmentation [18, 19].

In this paper we will discuss the behavior of transport characteristics for magnetic molecular-scale junctions in two response regimes: linear and nonlinear. In order to address the problem to a concrete physical situation, we will analyze the system consisting of carbon nanowires connected to ferromagnetic electrodes (Ni and Co). Since only delocalized π orbitals of organic molecules are involved in the conduction process, and Coulomb interactions can be important in determining transport properties of small systems, the molecule itself is described with the help of a Hückel model (π-electron approximation) with the electron interactions treated within the Hubbard approach. Coulomb interations within an atomic wire are treated by means of the Hartree-Fock (HF) approximation. The coupling to the electrodes is modeled through the use of a simplified broad-band theory.

**II. Computational scheme**

The Hamiltonian of the entire system of two electrodes spanned by a molecular wire can be expressed as a three-part sum: $H_{tot} = H_{el} + H_{mol} + H_{el-mol}$. The first term describes electrons in the electrodes:

$$H_{el} = \sum_{k,\sigma \in \alpha} \varepsilon_{k,\sigma} c^+_{k,\sigma} c_{k,\sigma}, \tag{1}$$

where $\alpha = L/R$ for the case of a left/right (source/drain) electrode, respectively. In the presence of bias voltage, one-electron energies $\varepsilon_{k,\sigma}$ are shifted in the following way: $\varepsilon_{k,\sigma} \to \varepsilon_{k,\sigma} + eV/2$ in the left electrode and $\varepsilon_{k,\sigma} \to \varepsilon_{k,\sigma} - eV/2$ in the right electrode. Chemical potentials of the electrodes are defined through the relations: $\mu_L = \varepsilon_F + eV/2$ and $\mu_R = \varepsilon_F - eV/2$ ($\varepsilon_F$ denotes the equilibrium Fermi level). The second term represents a linear N-atom chain, which is described within the Hubbard model approach [20, 21]:

$$H_{mol} = \sum_{i,j,\sigma}(\varepsilon_{i,\sigma}\delta_{i,j} - \beta)c^+_{i,\sigma}c_{j,\sigma} + U\sum_i n_{i\uparrow}n_{i\downarrow}, \quad (2)$$

where $\varepsilon_{i,\sigma}$ is the local site energy, $\beta$ is the hopping integral, $U$ is the on-site Coulomb interaction between two electrons with opposite spins, while $n_{i,\sigma}$, $c^+_{i,\sigma}$, and $c_{i,\sigma}$ denote the number, creation, and annihilation operators for an electron on site $i$ with spin $\sigma$. For the sake of simplicity we restrict the summation in Eq.2 to the simplest situation of nearest-neighbor atoms. By setting $U = 0$ in Eq.2, it reproduces the tight-binding (Hückel) Hamiltonian. Here we assume a uniform electric field between the electrodes and linear potential drop at the molecule (ramp model) [20-22]. Therefore, the local site energies $\varepsilon_{i,\sigma}$ are shifted due to this voltage ramp: $\varepsilon_{i,\sigma} \to \varepsilon_{i,\sigma} + eV[1 - 2i/(N+1)]/2$. The third term corresponds to the electron transfer from the electrodes onto the molecule:

$$H_{el-mol} = \sum_{k,\sigma\in\alpha;i} t_\alpha (c^+_{k,\sigma}c_{i,\sigma} + h.c.), \quad (3)$$

where $t_\alpha$ is the hopping integral responsible for the strength of the coupling with the $\alpha$ electrode. All the values of energy integrals ($\varepsilon$, $\beta$, $U$, $t$) are treated as parameters that can be modified within reasonable limits.

Further analyses are performed within the Hartree-Fock (HF) approximation, where the charge occupation numbers on particular sites are calculated using a self-consistent procedure. The HF problem is associated with a simplification of the molecular Hamiltonian (2), which can be rewritten in the form

$$H^{HF}_{mol} = \sum_{i,j,\sigma}(\bar{\varepsilon}_{i,\sigma}\delta_{i,j} - \beta)c^+_{i,\sigma}c_{j,\sigma}, \quad (4)$$

with the local site energy given by

$$\bar{\varepsilon}_{i,\sigma} = \varepsilon_{i,\sigma} + U\langle n_{i,\bar{\sigma}}\rangle. \quad (5)$$

The occupation number of the electrons on each site for particular voltages (nonequilibrium case) is determined self-consistently using the Keldysh formalism [23]:

$$\langle n_{i,\sigma}\rangle = -\frac{i}{2\pi}\int_{-\infty}^{+\infty} d\omega G^<_{i\sigma,i\sigma}(\omega). \quad (6)$$

The lesser Green function $G^<$ can be obtained from the Dyson equation and expressed in the general form as

$$G^<_{i\sigma,j\sigma} = \sum_{i',j'} G^r_{i\sigma,i'\sigma}\Sigma^<_{i'\sigma,j'\sigma}G^a_{j'\sigma,j\sigma}. \quad (7)$$

The superscripts $r$ and $a$ denote the retarded and advanced Green functions, respectively:

$$G^r(\omega) = [J\omega - H^{HF}_{mol} - \Sigma^r]^{-1} \quad (8)$$

and $G^a = [G^r]^*$ (Here $J$ denotes the unit matrix of the dimension equal to the molecular Hamiltonian $N \times N$). Since the molecule is contacted with the electrodes only through the atoms at the ends of the wire, the lesser self-energy can be written as follows:

$$\Sigma^{<}_{i\sigma,j\sigma}(\omega) = 2i\delta_{i,j}\left[\delta_{i,1}\Delta_{L\sigma}(\omega)f_L(\omega) + \delta_{i,N}\Delta_{R\sigma}(\omega)f_R(\omega)\right], \tag{9}$$

where $f_\alpha$ is the Fermi distribution function in the $\alpha$ electrode. Furthermore, the retarded and advanced self-energy functions are given by

$$\Sigma^{r}_{i\sigma,j\sigma}(\omega) = \delta_{i,j}\left[\delta_{i,1}[\Lambda_{L\sigma}(\omega) - i\Delta_{L\sigma}(\omega)] + \delta_{i,N}[\Lambda_{R\sigma}(\omega) - i\Delta_{R\sigma}(\omega)]\right] \tag{10}$$

and $\Sigma^a = [\Sigma^r]^*$. The real and imaginary terms of the self-energy components are not independent from each other, being related through the Hilbert transform [20]:

$$\Lambda_{\alpha\sigma}(\omega) = P\int dz \frac{\Delta_{\alpha\sigma}(z)}{[\pi(\omega - z)]}, \tag{11}$$

where $P$ is the Cauchy principal value. In order to speed up the computations we approximate the contact self-energies with the help of their imaginary elements only (neglecting their real parts as responsible for energy shifts). In our case

$$\Delta_{\alpha\sigma}(\omega) = \pi\rho_{\alpha\sigma}t_\alpha^2, \tag{12}$$

where $\rho_{\alpha\sigma}$ is the local density of states (at the Fermi energy level) for electrons with spin $\sigma$ in the $\alpha$ electrode. This assumption means that the local density of states in both electrodes is constant over an energy bandwidth and zero otherwise. The dispersionless coupling parameters are commonly used in the literature and are usually sufficient in describing broad-band metals [11, 12, 22, 24, 25]. The model here indicates that spin-polarization of the junction is determined only through the DOS of ferromagnets, but the realistic situation is much more complicated [26].

The current flowing through the device can be computed from the time evolution of the occupation number for electrons in the left (or equivalently right) electrode $N_\alpha = \sum_{k,\sigma \in \alpha} c^+_{k,\sigma}c_{k,\sigma}$ and can be expressed by the lesser Green function [23, 27]:

$$I(V) = -e\frac{d}{dt}\langle N_\alpha \rangle = \frac{e}{h}\sum_{i,k,\sigma}t_\alpha \int_{-\infty}^{+\infty} d\omega \left[G^{<}_{i\sigma,k\sigma}(\omega) + c.c.\right]. \tag{13}$$

After applying the Dyson equation, the current formula can be written with the help of the retarded Green function:

$$I(V) = \frac{4e}{h}\sum_\sigma \int_{-\infty}^{+\infty} d\omega [f_L(\omega) - f_R(\omega)]\Delta_{L\sigma}(\omega)\Delta_{R\sigma}(\omega)|G^{r}_{1\sigma,N\sigma}(\omega)|^2. \tag{14}$$

Conductance is then given as a derivative of the current with respect to voltage: $G(V) = dI(V)/dV$, while tunnel magnetoresitance (TMR coefficient) is computed as

$$TMR(V) = \frac{G_P(V) - G_{AP}(V)}{G_P(V)}. \tag{15}$$

In our calculations we have assumed that the transport process is purely coherent and elastic. It means that the current conservation rule is fulfilled on each site and for any energy $\omega$. Furthermore, the method presented in this work (based on the HF approximation) neglects all the many-body effects and electronic correlations.

## III. Numerical results and comments

As an example we consider the model of a linear carbon-atom chain connected to two ferromagnetic electrodes. The molecular description itself includes only π-electrons of hydrocarbons (and is based on the assumption of one $2p_z$-basis function for each carbon atom), while the coupling to the electrodes is treated within a broad-band theory. This is a test case simple enough to analyze all the essential physics in detail. In order to simulate conjugated molecules, we choose the following energy parameters (given in eV) [20]: $\varepsilon = 0$ (the reference energy), $\beta = 2.4$, $U = 2$. In this work we assume realistically that the Fermi level is fixed exactly in the middle of the DOS spectra of the molecule ($\varepsilon_F = 0$). Since ferromagnets have unequal spin-up and spin-down populations, their densities of states for both spin orientations are different. Here we adopt such densities for Ni and Co electrodes from the work of Babiaczyk and Bułka [11] as obtained from band-structure calculations performed using the tight-binding version of the linear muffin-tin orbital method in the atomic sphere approximation: $\rho_{Ni\uparrow} = 0.1897$, $\rho_{Ni\downarrow} = 1.7261$, $\rho_{Co\uparrow} = 0.1740$, $\rho_{Co\downarrow} = 0.7349$. Assuming symmetric coupling at both ends of the molecule and setting typical hopping parameters as $t_L = t_R = 0.5$, we obtain the following self-energy terms: $\Delta_{Ni\uparrow} = 0.1490$, $\Delta_{Ni\downarrow} = 1.3557$, $\Delta_{Co\uparrow} = 0.1367$, $\Delta_{Co\downarrow} = 0.5772$. The temperature energy of the system is assumed to be equal to that of room temperature, $k_B T = 0.025$ (However, the results are not particularly sensitive to temperature).

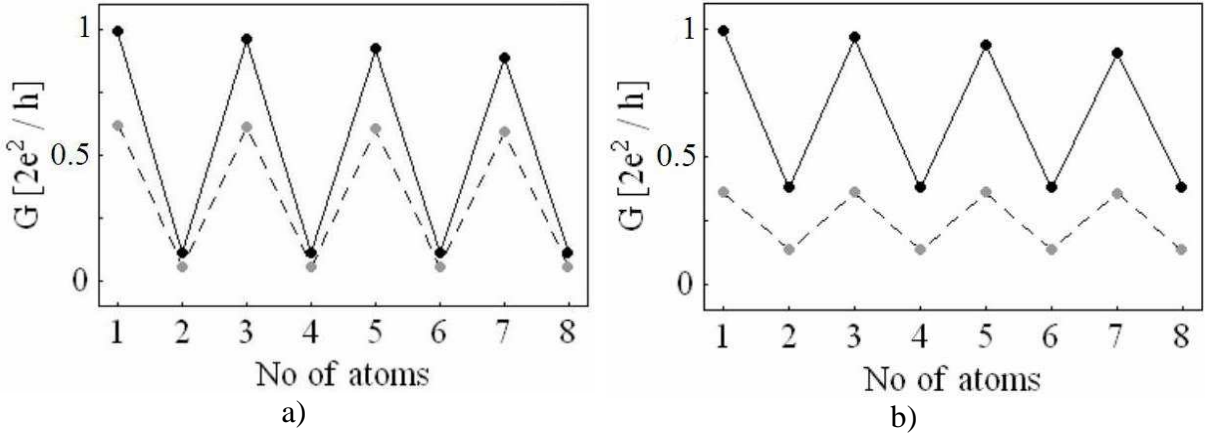

a)      b)

Figure 1: Linear conductance as a function of the number of carbon atoms attached to a) Co electrodes and b) Ni electrodes. Solid curves (black circles) and broken curves (grey circles) correspond to parallel and antiparallel of the electrodes' magnetization, respectively. The other parameters of the model (given in eV): $\varepsilon = 0$, $\beta = 2.4$, $U = 2$, $t_L = t_R = 0.5$, $k_B T = 0.025$.

### A. Linear response regime

Now we discuss the length behavior of transport characteristics found in the linear response regime. It is clear that at low bias voltages ($V \leq 0.1$ Volt), the current becomes a linear function of an applied bias: $I = GV$. The method here allows us to compute the zero-bias conductance using the Landauer-type expression [28]:

$$G = \frac{4e^2}{h} \sum_\sigma \int_{-\infty}^{+\infty} d\omega F_T(\omega) \Delta_{L\sigma}(\omega) \Delta_{R\sigma}(\omega) \left| G^r_{1\sigma,N\sigma}(\omega) \right|^2, \tag{16}$$

where the thermal broadening function is given through the relation [22, 29, 30]:

$$F_T(\omega) \equiv -\frac{\partial f(\omega)}{\partial \omega} = \frac{1}{4k_B T} \mathrm{sech}^2\left[\frac{\omega - \varepsilon_F}{2k_B T}\right]. \tag{17}$$

Figure 1 shows the linear conductance as a function of the number of carbon atoms attached to ferromagnetic electrodes (Ni and Co). Here conductance for parallel alignment of magnetizations in the electrodes reaches higher values than for the case of antiparallel alignment. However, conductance oscillations with increasing length of the wire are also observed for both cases, with odd-numbered nanowires showing a higher conductance than even-numbered ones. But in our studies conductance values fall in a range of 0 to $G_0$ (maximal value of conductance for one spin-degenerated level). The amplitude of the oscillations strongly depends on the strength of the molecule-to-electrodes coupling and the position of the Fermi level in relation to the electronic structure of the atomic wire. Remember that molecular energy levels constitute discrete conducting channels between two reservoirs of charge carriers.

The origin of oscillatory conductance is very simple. For an even number of carbon atoms, there are occupied bonding states and unoccupied antibonding states separated by the HOMO-LUMO gap. Since we chose the Fermi level to be fixed exactly in the middle of the molecular DOS spectrum, electronic transport is in the off-resonance regime, that is, transmission probability is much smaller than unity. However, for an odd number of carbon atoms a half-filled nonbonding state exists in the middle of molecular DOS, and consequently we have an open channel for conductance – a resonance regime, where transmission probability is close to unity. So periodic changes of the conductance with increasing length of the nanowire are associated with switching between resonance and off-resonance transport for odd- and even-numbered wires.

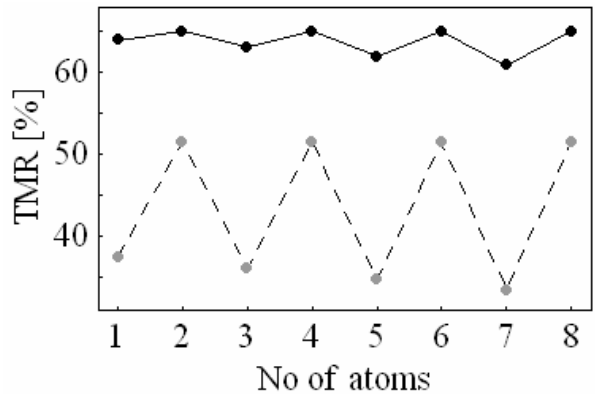

Figure 2: Tunnel magnetoresistance as a function of the number of carbon atoms attached to Ni electrodes (solid curve, black circles) and Co electrodes (broken curve, grey circles), respectively. The other parameters of the model are the same as in Fig.1.

In Fig.2 we plot tunnel magnetoresistance as a function of the number of carbon atoms attached to ferromagnetic electrodes (Ni and Co) in the linear response regime. The TMR coefficient reaches higher values for Ni electrodes in comparison with Co electrodes.

Moreover, magnetoresistance varies periodically with increasing length of the atomic wire, which is a straightforward consequence of conductance oscillations. However, odd-numbered nanowires show a lower value of TMR coefficient than even-numbered ones, in opposition to conductance behavior (TMR oscillations are out of phase with conductance ones.). Oscillations for Ni electrodes are not so evident as for Co electrodes. This effect again strongly depends on the strength of the molecule-to-electrodes coupling and the position of the Fermi level in relation to the electronic structure of the molecular wire. Anyway, the choice of Coulomb integral has a negligibly small influence on our zero-bias results, since the value of the $U$-parameter is essential only for finite voltages (as will be discussed later in this paper).

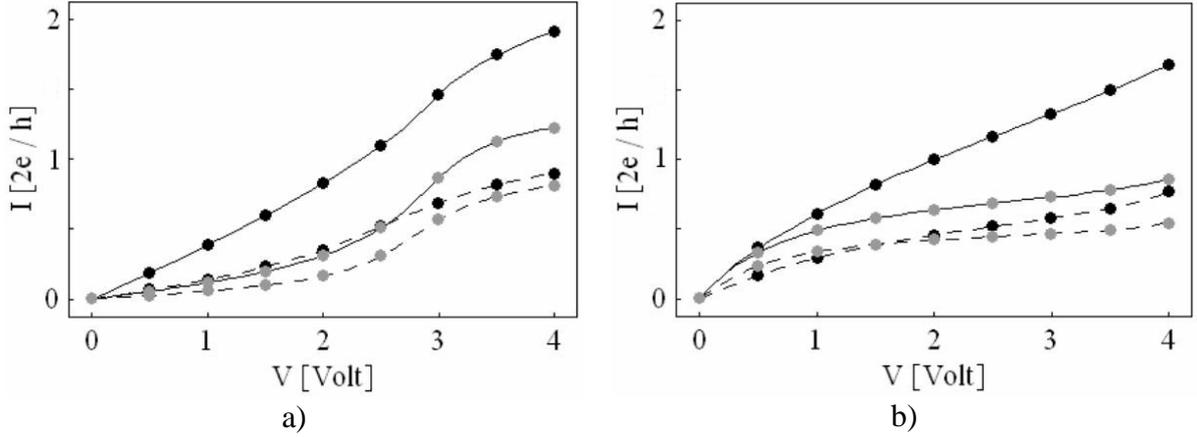

Figure 3: Current-voltage characteristics for the device made of four carbon atoms (a: $N = 4$) and five carbon atoms (b: $N = 5$) attached to Ni electrodes (black circles) and Co electrodes (grey circles), respectively. Solid and broken curves correspond to parallel and antiparallel alignment of the electrodes' magnetization. The other parameters of the model are the same as in Fig.1.

### B. Nonlinear response regime

Now we discuss the voltage dependence of transport characteristics found in the nonlinear response regime. For higher voltages ($V > 0.1$ Volt), the current becomes a nonlinear function of an applied bias (see Eq.13). Such nonlinearity arises because of an exponential dependence of Fermi functions on bias voltage and variation of the molecular Green function, due to the voltage shift of site energy levels in the wire. Figure 3 shows I-V curves for the case of four-atom and five-atom nanowires connected to ferromagnetic electrodes (Ni and Co). Such I-V curves are fairly smooth, since the energy level broadening due to its contact with the electrodes is significant. The current for the P alignment reaches higher values than for the AP configuration in the case of the analyzed materials. Similarly, the current for Ni-based junctions reaches higher values in comparison with Co-based junctions.

Although the predicted order of the magnitude of the current values (hundreds of μA) is comparable with some *ab initio* computations [24, 33, 34], the discrepancy with the experimental results indicates that the coupling of the molecule to the electrodes can be smaller than estimated. There are few factors that can be crucial in determining the parameter of the coupling strength: the atomic-scale contact geometry; the nature of the molecule-to-electrodes coupling (chemisorption or physisorption); or even the variations of surface properties due to adsorption of a molecular monolayer. Additional effects that can alter the value of the current flowing through the junction are associated with some temperature effects (hot electrons and vibrational coupling) or local disorder in the electrodes near the contacts (electron localization) [35].

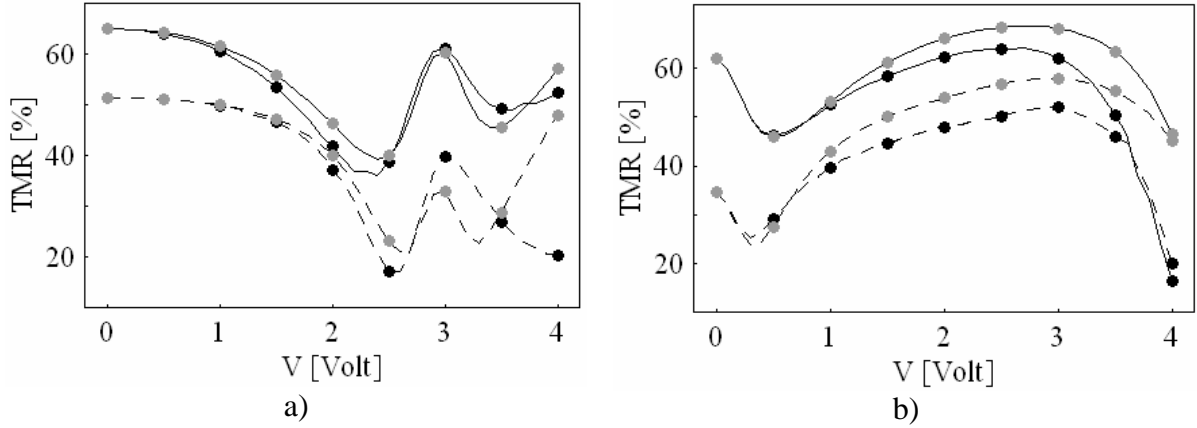

Figure 4: Tunnel magnetoresistance for the device made of four carbon atoms (a: $N=4$) and five carbon atoms (b: $N=5$) attached to Ni electrodes (solid curves) and Co electrodes (broken curves), respectively. Grey and black circles correspond to different Hubbard parameters $U=0$ and $U=2$. The other parameters of the model are the same as in Fig.1.

In Fig.4 we plot tunnel magnetoresistance for the devices made of four-atom and five-atom nanowires connected to ferromagnetic electrodes (Ni and Co). Here we can observe oscillations of the TMR coefficient with increasing bias voltage. Such an effect is independent of the strength of the $U$-parameter (and even the presence or absence of potential drop along the molecular wire). It turned out that Coulomb repulsion is important in determining magnetoresistance (or, alternatively, conductance) only in the case of higher voltages (compare grey and black circles in Fig.4). The effect of electron-electron interactions (taken into account within the molecular system) is mostly to reduce the TMR parameter.

## IV. A brief summary

Spin-dependent transport calculations were performed for molecular-scale devices made of atomic nanowires attached to ferromagnetic electrodes. The molecular system was treated as a linear Hubbard chain (at the Hartree-Fock level), while the coupling to the electrodes was described within a broad-band theory. Parameters of the model were chosen in such a way to simulate carbon-atom wires connected to Ni and Co ferromagnets. First of all it was shown that magnetoresistance can be large (tens of percent), varies periodically with increasing length of atomic wire (in the linear response regime), and oscillates with increasing of bias voltage (in the nonlinear response regime).

This work brings us nearer to understanding the electrical conduction on a molecular scale and to determining the main factors that control transport through molecular junctions. Our predictions could help in designing future electronic nanocircuits. Since the TMR effect for the analyzed structures is predicted to be large, these junctions could play the role of the simplest switches, where the switching mechanism between two states of low and high conductance is associated with application of an external magnetic field (which is needed in order to change magnetization in one of two electrodes). Maybe one day it will be possible to connect individual molecular-scale devices into a properly working integrated circuit (IC) and construct small supercomputers with extraordinary parameters.

**Acknowledgments**

K. Walczak is very grateful to B. Bułka and T. Kostyrko for many valuable discussions. This work was supported by the European Commission contract No. HPRN-CT-2002-00282.